\documentclass[pra,aps,twocolumn,showpacs,nofootinbib]{revtex4}

\usepackage{amsfonts,amsmath,bm,amsthm,amssymb}
\usepackage[OT4]{fontenc}
\usepackage{color}
\usepackage{graphicx}
\usepackage{epstopdf}
\usepackage{verbatim}

\usepackage{hyperref}

\hypersetup{
   colorlinks=true,         
   linkcolor=blue,          
   citecolor=blue,        
   urlcolor=Violet            
}

\usepackage{bookmark}

\newcommand{\bra}[1]{\ensuremath{\left\langle#1\right|}}
\newcommand{\ket}[1]{\ensuremath{\left|#1\right\rangle}}
\newcommand{\braket}[2]{\ensuremath{\left\langle#1 \vphantom{#2}\middle|  #2 \vphantom{#1}\right\rangle}}
\newcommand{\ketbra}[2]{\ensuremath{\left|#1\right\rangle\left\langle#2\right|}}
\newcommand{\matrixel}[3]{\ensuremath{\left\langle #1 \vphantom{#2#3} \right| #2 \left| #3 \vphantom{#1#2} \right\rangle}}
\newcommand{\tr}[1]{\mathrm{Tr}\left( #1 \right)}
\newcommand{\iden}{\mathbb{I}}

\renewcommand{\v}[1]{\ensuremath{\underline{\boldsymbol #1}}}

\def\>{\rangle}
\def\<{\langle}

\def\D{ {\cal D} }

\newcommand{\be}{\begin{equation}}
\newcommand{\ee}{\end{equation}}

\newcounter{thm}
\newtheorem{maximal_quantum_u}{Lemma}
\newtheorem{trivial_bound}[thm]{Theorem}
\newtheorem{non_linearity}[thm]{Theorem}

\begin{document}

\title{Quantum and classical entropic uncertainty relations}

\author{Kamil Korzekwa, Matteo Lostaglio, David Jennings, Terry Rudolph}
\affiliation{Department of Physics, Imperial College London, London SW7 2AZ, United Kingdom}

\begin{abstract}

How much of the uncertainty in predicting measurement outcomes for noncommuting quantum observables is genuinely quantum mechanical? We provide a natural decomposition of the total entropic uncertainty of two non-commuting observables into a classical component, and an intrinsically quantum mechanical component. We show that the total quantum component in a state is never lower or upper bounded by any state-independent quantities, but instead admits ``purity-based'' lower bounds that generalize entropic formulations such as the Maassen-Uffink relation. These relations reveal a non-trivial interplay between quantum and classical randomness in any finite-dimensional state.

\end{abstract}

\pacs{03.65.Ta, 89.70.Cf, 03.67.Mn}

\maketitle

\section{Introduction}
\label{sec:intro}

Quantum phenomena are notoriously unpredictable. While classical uncertainty arises from ignorance, quantum uncertainty is hardwired in, such that even for a single fixed measurement and a pure quantum state of maximal knowledge we typically can only make probabilistic predictions. The situation worsens when we consider two potential measurements of a system that do not commute -- there exist fundamental constraints on our ability to make predictions about either possible set of outcomes. The most celebrated such constraint is the Heisenberg-Robertson uncertainty relation \cite{robertson1929uncertainty}, which bounds the product of the variances in the two possible measurement outcomes in terms of the expectation of the commutator of the observables. A weakness of this formulation is that it typically depends on the particular quantum state of the system to be measured, and as such the bound can be trivial if the state lies in the kernel of the commutator.

A different approach, pioneered by Deutsch  \cite{deutsch1983uncertainty} and strengthened by Maassen and Uffink \cite{maassen1988generalized}, is to lower bound in a state-independent manner the sum of the Shannon entropies of the measurement outcome probability distributions. Such an entropic uncertainty relation (EUR) is particularly interesting within the context of mutually unbiased measurement bases (MUBs) (see Ref. \cite{wehner2010entropic} and references therein for extensions and generalizations).

Quantum states are typically mixed, which means that some of our inability to make perfect predictions is actually classical. The original EUR does not separate the uncertainty that arises from this classical ignorance and the ``genuinely quantum'' uncertainty. Our goal here is to do such a separation, something considered for Heisenberg-Robertson-type uncertainty relations in Ref. \cite{luo2005heisenberg}, where it was shown that quantum-only uncertainties satisfy a similar bound to the full uncertainty. Here we will show that this is not possible for additive decompositions of EURs into quantum and classical parts, if we wish to retain the state-independence of the bound. However, we show that splitting the uncertainty can be done, and a bound on the sum of the quantum uncertainties achieved, if we impose only the weak constraint that the purity of the set of states under question is fixed. This leads us to introduce ``purity-based bounds.'' In this paper we develop certain purity-based entropic bounds and compare them with recently introduced majorization bounds. We also show that nonlinearity in $S(\rho)$ is a general feature of strong purity-based bounds. Finally, we analyze the structure of states minimizing the total uncertainty for fixed purity and show that it displays non-trivial structure.

\section{Entropic Uncertainty Relations}

The basic idea underlying entropic uncertainty relations is extremely simple. Given a nondegenerate observable $O$ acting on a $d$-dimensional Hilbert space,
\be
\label{eq:observable}
O=\sum_{i=1}^do_i \ketbra{o_i}{o_i},
\ee
the projective measurement of $O$ in the state $\rho$ gives rise to a distribution $p_i(O,\rho)=\tr{\ketbra{o_i}{o_i}\rho}$. The entropic measure of uncertainty about $O$ in a state $\rho$ is then the Shannon entropy of the distribution,
\be
\label{eq:shannon}
H_O(\rho):=-\sum_{i=1}^dp_i(O,\rho)\ln \left(p_i(O,\rho)\right),
\ee
whose use as a measure of uncertainty is well established in classical and quantum information theory. 

Given two, possibly noncommuting observables $A$ and $B$ with eigenbases $\{\ket{a_i}\}$ and $\{\ket{b_j}\}$, we ask: is there a lower bound to the combined uncertainty $H_A(\rho) + H_B(\rho)$ in the state $\rho$? A definitive answer was first given by Deutsch \cite{deutsch1983uncertainty}, then later strengthened to provide us the celebrated Maassen-Uffink relation \cite{maassen1988generalized},
\begin{eqnarray}
\label{eq:muEUR}
H_A(\rho) + H_B(\rho) \ge - 2\ln c_{AB},
\end{eqnarray}
where $c_{AB}= \max_{ij} |\braket{a_i}{b_j}|$ yields the state-independent lower bound.

This relation has been improved in many respects. A tighter relation for observables fulfilling $c_{AB}>1/\sqrt{2}$ has been found using the Landau-Pollak uncertainty relation \cite{devicente2008improved}. The study of the entropic uncertainty relations in the presence of classical or quantum side information brought other improvements \cite{berta2010uncertainty,coles2014improved}. In particular, the case of a trivial memory gives a bound on $H_A(\rho) + H_B(\rho)$ in terms of the von Neumann entropy of the state \cite{berta2010uncertainty}. Very recently a majorization approach \cite{puchala2013majorization,friedland2013universal} lead to improvements based on a more fine-grained measure of overlap between observables, where one looks at all $|\braket{a_i}{b_j}|$ \cite{rudnicki2014strong}.

\section{Splitting Total Uncertainty into Quantum and Classical Parts}
\label{sec:splitting}

While intuitive and operationally meaningful, the entropic measure $H_O(\rho)$ quantifies total uncertainty and, as such, does not carry information about its origin (classical or quantum).  For example, in the $d=2$ case the Pauli observable $Z$ has maximal uncertainty $H_Z= \ln 2$ in both the states $|+\>\<+|$ and $ \iden/2$. However, it is evident that in the former case the uncertainty is entirely quantum mechanical, whereas in the latter it is entirely classical. As such, the entropic measure of the sum of the Shannon entropies provides only a blunt quantification of quantum uncertainties that deserves a finer analysis. The idea of a classical-quantum decomposition has been also applied to measures of correlations, see, e.g., Refs. \cite{ollivier2001quantum, henderson2001classical, luo2008quantum}. We will shortly see how our framework can be related to some of these results.

\subsection{The Luo criteria for measurement uncertainties}

Obviously there is no single correct way to decompose the total measurement uncertainty of the observable $O$ in a state $\rho$ into quantum and classical components. However it is quite straightforward to establish natural criteria that such a decomposition should obey. Such a set of conditions was recently formulated by Luo \cite{luo2005quantum}, and represents the minimal conditions that any quantum uncertainty measure $Q(O,\rho)$ and any classical uncertainty measure $C(O,\rho)$ should satisfy. Specifically, they demand that:

\begin{enumerate}
\item \label{luo1} If a state $\rho$ is pure, then $C(O,\rho)$ should vanish.
\item \label{luo2} If $[\rho,O]=0$, then the state is diagonal in the eigenbasis of the observable $O$ and so $Q(O,\rho)$ should vanish.
\item Classical mixing increases the classical, but not the quantum, uncertainty, and so $Q(O,\cdot)$ should be convex and $C(O,\cdot)$ should be concave in their second arguments.
\end{enumerate}
To this criteria we add further conditions specific to an entropic scenario:
\begin{enumerate}
\setcounter{enumi}{3}
\item $0 \leq Q(O,\rho), C(O,\rho) \leq H_O(\rho)$.
\item $Q(O,\cdot)$ and $C(O,\cdot)$ are functions of the probability distribution over the measurement outcomes of observable $O$ and not of its eigenvalues.
\end{enumerate}

The relative merits or weaknesses of these conditions can certainly be debated, but in what follows we simply use them as a guide for our entropic decomposition and leave extensions to future work.

\subsection{Classical and quantum decomposition for entropic uncertainty relations.} 

We wish to develop a meaningful decomposition into classical and quantum entropic measures of uncertainty suitable for extending the Maassen-Uffink relation, which respects the Luo criteria.
In light of these criteria, we observe that the central measurement entropy for nondegenerate observables can be expressed as \mbox{$H_O(\rho)=S(\D_O (\rho))$}, where \mbox{$S(\rho) = -\mathrm{Tr}[ \rho \ln \rho]$} is the von Neumann entropy of a quantum state $\rho$, and $\D_O(\cdot)$ is the dephasing map,
\be
\mathcal{D}_O(\rho)=\sum_{i=1}^d\matrixel{o_i}{\rho}{o_i}\ketbra{o_i}{o_i},
\ee
which sends to zero all the off-diagonal terms when the density matrix is written in the eigenbasis of $O$ (an extension to degenerate observables can be found in Appendix A). The projective measurement of $O$ is a repeatable measurement, and so it is reasonable to demand that a second measurement of $O$ (in which the prior classical outcome of the first measurement is discarded) should not reveal any quantum uncertainty in the state, and be entirely classical.

In light of this, we take $Q(O,\rho) : = S(\rho || \D_O (\rho))$ as the appropriate measure of quantum uncertainty for the measurement of $O$ in the state $\rho$.  The following geometrical characterization of $Q$ as a ``distance'' from the manifold of classical states further justifies our definition: 
\be
\label{eq:geo_quant}
Q(O,\rho) := S(\rho || \D_O (\rho)) = \min_{\sigma \in \mathcal{I}} S(\rho||\sigma),
\ee
where $\mathcal{I}$ is the set of states diagonal in the eigenbasis of $O$ (this is a direct application of Theorem 4 of Ref. \onlinecite{gour2009measuring}).  Moreover, if we take $C(O,\rho) := S(\rho)$ as our measure of classical uncertainty \cite{luo2009relative}, we obtain an \emph{additive} decomposition of the total entropic uncertainty,
\be
\label{eq:splitting}
H_O(\rho) = Q(O,\rho) + C(O,\rho).
\ee
To prove the equality in Eq. \eqref{eq:splitting}, we note that
\begin{eqnarray*}
-\tr{\rho\ln[\D_O(\rho)]}&=&-\tr{\rho \D_O\{\ln[\D_O(\rho)]\}}\\
&=&-\tr{\D_O(\rho)\ln[\D_O(\rho)]}=S(\D_O(\rho)),
\end{eqnarray*}
so that
\begin{eqnarray*}
Q(O,\rho)=S(\rho || \D_O (\rho))&=&-S(\rho)-\tr{\rho\ln[\D_O(\rho)]}\\
&=&-S(\rho)+S(\D_O(\rho)).
\end{eqnarray*}

It is relatively straightforward to check that these measures conform to the Luo criteria, using well-known properties of the von Neumann entropy. Convexity of $Q$ follows immediately from the joint convexity of relative entropy. 

In addition to providing an additive decomposition, our choice of quantum uncertainty $Q(O,\rho)$ has a natural interpretation as a measure of the superpositions present in $\rho$ with respect to the eigenbasis of $O$ \cite{aberg2006superposition}, and more recently has provided monotones within the resource theories of coherence \cite{plenio2013coherence} and $U(1)$-asymmetry \cite{gour2008resource}. Moreover, within the quantum memory approach our choice of $Q(O,\rho)$ corresponds to quantum side information introduced by the system $E$ purifying $\rho$; i.e., it is equivalent to conditional entropy $H(O|E)$, as discussed in Ref. \cite{coles2012uncertainty}. $Q$ can be also related to discord-like quantities. In a bipartite system, the minimum of $Q$ over all maximally informative local observables \cite{girolami2013characterizing} is equal to the von Neumann entropy of entanglement $E(\rho_{12}):=S(\mathrm{Tr}_2(\rho_{12}))$:
\be
\nonumber
Q^{(1)} (\ket{\psi}_{12}\bra{\psi}) := \min_{O_1} Q(O_1\otimes \iden_2,\ket{\psi}_{12}\bra{\psi} ) = E(\rho_{12}).
\ee
Indeed, $Q^{(1)}$ is known as thermal or one-way discord \cite{modi2012discord}. These connections make the measure $Q$ additionally attractive and facilitate interpretation within a broader framework.

Note that our measure of classical uncertainty does not depend on the choice of the observable. This is due to the fact that we are considering here nondegenerate observables, projective measurements of which are perfectly sharp. In this situation any classical uncertainty can only be due to the fact that we are sampling a mixed state. A similar situation would be true for perfectly sharp measurements in classical physics. However, similarly to coarse-grained measurements in classical physics, the projective measurements of degenerate observables in quantum physics can be the source of classical uncertainty dependent on the degeneracy. We analyze this extension in Appendix A.

Whereas the Maassen-Uffink relation bounds the total entropic uncertainty $H_A + H_B$, we would now like to establish a finer set of conditions on the quantum component of the total uncertainty. However before we do this we show, using Luo's criteria, that $Q(A,\cdot)+Q(B,\cdot)$ is entirely unconstrained over the set of all states, i.e., no state-independent lower (or upper) bound is possible for the total quantum uncertainty of $A$ and $B$ in the state $\rho$.

\subsection{No non-trivial state-independent bound for quantum or classical uncertainties}

While we are interested in a particular quantum-classical splitting, the following argument applies more generally.  In Ref. \cite{friedland2013universal}, Friedland \textit{et al.} characterized the most general \emph{uncertainty function}, i.e., the most general $U:\bf{p} \longmapsto \mathbb{R}^+$ invariant under relabellings of the probability vector $\bf{p}$ and monotonically increasing under \emph{random relabellings}.\footnote{For every $q \in [0,1]$ and every permutation $\pi$, \mbox{$U({\bf p}) \leq U(q {\bf p} + (1-q)\pi {\bf p})$}.} In other words, $U$ is required to preserve the majorization ordering and is then a Schur-concave function. 

We start by giving a general result on the total uncertainty of the measurement outcomes of two observables. Let $\mathbf{p}(A,\rho)$ and $\mathbf{p}(B,\rho)$ denote vectors of probability outcomes of $A$ and $B$, respectively, in state $\rho$ and \mbox{$U_{max}=\max_{\mathbf{p}}U(\mathbf{p})$}. Then, the following lemma holds,
\begin{maximal_quantum_u}
\label{th:maximal_quantum_u}
For every pair of nondegenerate observables $A$ and $B$ there exists a pure state $\ket{\psi^*}$ that simultaneously maximizes the total uncertainty of both observables, 
\begin{equation*}
U(\mathbf{p}(A,\ket{\psi^*}))=U(\mathbf{p}(B,\ket{\psi^*}))=U_{max}.
\end{equation*}
\end{maximal_quantum_u}
\begin{proof}
For every pair of observables $A$, $B$ there exists a pure state $\ket{\psi^*}$ which is unbiased in eigenbases of both observables \cite{korzekwa2013operational}: 
\begin{equation*}
\mathbf{p}\left(A,\ket{\psi^*}\right) = \mathbf{p}\left(B,\ket{\psi^*}\right)=  \{1/d,...,1/d\}.
\end{equation*}
The uniform distribution, $\{1/d,...,1/d\}$, is majorized by all other distributions. Hence, as $U$ is Schur-concave, one gets
\begin{equation*}
\forall \mathbf{q}\quad U(\mathbf{q})\leq U(\mathbf{p}(A,\ket{\psi^*}) = U(\mathbf{p}(B,\ket{\psi^*}) := U_{max}.
\end{equation*}
\end{proof}

Suppose we now want to split the general measure of total uncertainty $U$ into the sum of classical and quantum uncertainty components,
\be
\label{eq:sumsplitting}
U=Q+C,
\ee
where we have two non-negative real-valued functions \mbox{$Q:(A,\rho) \mapsto \mathbb{R}^+$}, the \emph{quantum uncertainty}, and \mbox{$C:(A,\rho) \mapsto \mathbb{R}^+$}, the \emph{classical uncertainty}. Given the additive splitting, defined by Eq. \eqref{eq:sumsplitting}, we could wonder if we can find a state-independent upper or lower bound on \mbox{$Q(A,\cdot)+Q(B,\cdot)$} or \mbox{$C(A,\cdot)+C(B,\cdot)$} only. From the previous lemma we immediately infer this is impossible, if one demands the Luo criteria of the quantum and classical uncertainties. Specifically we have the following,
\begin{trivial_bound}
\label{th:trivialbound}
No additive splitting admits a nontrivial state-independent bound for \mbox{$Q(A,\cdot)+Q(B,\cdot)$} or \mbox{$C(A,\cdot)+C(B,\cdot)$} if Luo's criteria \ref{luo1} and \ref{luo2} are satisfied. In other words there are no $A$, $B$, $c(A,B)>0$ and $d(A,B)<2U_{max}$ such that:
\be
\nonumber
 \forall \rho\quad c(A,B)< Q(A,\rho)+Q(B,\rho) < d(A,B).
\ee
\end{trivial_bound}
\begin{proof}
Let us fix general nondegenerate observables $A$ and $B$.  From Theorem \ref{th:maximal_quantum_u} there is always a pure state $\ket{\psi^*}$ achieving \mbox{$U(\mathbf{p}(A,\ket{\psi^*}) = U(\mathbf{p}(B,\ket{\psi^*}) = U_{max}$}. But from Luo criterion \ref{luo1}, $C$ vanishes on pure states, so
\be
\nonumber
Q(A,\ket{\psi^*}) + Q(B,\ket{\psi^*})=2U_{max}.
\ee
The maximally mixed state is diagonal in any basis; therefore, from Luo criterion \ref{luo2},
\be
\nonumber
Q(A,\iden/d) + Q(B,\iden/d) =0.
\ee
Given the additive splitting and that the sum of $Q$'s is unconstrained, then we also deduce that there is also no constraint on the sum of classical components.
\end{proof}
\noindent In particular this implies that, in the case of EUR and the splitting proposed in this paper,
\be
\nonumber
0\leq Q(A,\rho) + Q(B,\rho) \leq 2\ln d,
\ee
and only the total uncertainty has a state-independent lower bound. To account for this we instead develop bounds that are conditioned on fixed values of classical measurement uncertainty. For our choice of classical and quantum uncertainties, this will lead us to a refinement of EUR in terms of ``purity-based" lower bounds.

\section{Quantum uncertainty relations and purity-based lower bounds}

We now establish concrete lower bounds on the total quantum uncertainty for the measurement of two observables $A$ and $B$ in a state $\rho$. As discussed, we seek bounds that are conditional on the degree of \emph{classical} uncertainty in the state. Schematically, we would like to obtain entropic relations for the total quantum uncertainty of the form
\begin{equation}
Q(A,\rho) + Q(B,\rho) \ge f(A,B, \mbox{``purity of } \rho \mbox{''}).
\end{equation}
In this work we focus on the case where $Q(A,\cdot)$ and $Q(B,\cdot)$ are the entropic quantum uncertainties of Eq. \eqref{eq:geo_quant} and the purity is measured by the von Neumann entropy,
\begin{equation}
\label{eq:boundgeneralform}
Q(A,\rho) + Q(B,\rho) \ge f(A,B, S(\rho)). 
\end{equation}
We may additionally require any purity-based bound to satisfy the following two desiderata:
\begin{enumerate}
\item \label{desiderata1} Being at least as strong as the Maassen-Uffink bound:
\be
\nonumber
\forall \rho, \; \; \forall A,B, \; \mbox{Eq.} \eqref{eq:boundgeneralform} \Rightarrow H_{A}(\rho) + H_{B}(\rho) \geq -2 \log c_{AB}.
\ee
\item For $d$-dimensional space $f(A,B, \ln d) =0$.\label{desiderata2}
\end{enumerate}
The second desideratum captures the classical feature of the maximally mixed state, namely that it should not exhibit any quantum uncertainty, consistently with the vanishing of all coherences. We will refer to purity-based bounds satisfying properties \ref{desiderata1} and \ref{desiderata2} as \emph{strong purity-based bounds} (SPB). 

In what follows we derive several relations having the form of Eq. \eqref{eq:boundgeneralform}. However, the following general restriction on the nonlinear nature of all strong purity-based bounds holds:
\begin{non_linearity}[No linear SPB]
\label{th:nolinearbound}
For all $d >2$ all strong purity-based bounds are nonlinear in $S(\rho)$.
\end{non_linearity}
\noindent The proof is based on direct construction of a counterexample for the weakest SPB and is given in Appendix B. Notice that since \mbox{$H_{O}(\rho) = Q(O,\rho) + S(\rho)$}, the same conclusion applies to the usual entropic uncertainty relations. As all bounds proposed so far in the literature are, to our knowledge, linear in $S$, they are inevitably either weaker than Maassen-Uffink for at least some states and observables, or they are not tight for the maximally mixed state. 

\subsection{Mutually unbiased observables}

Since the quantum uncertainty measure is the relative entropy between the state $\rho$ and its dephased output state following the projective measurements, we can make use of certain well-known entropic properties to develop meaningful lower bounds on the total quantum uncertainty in a state. This approach is similar to the one used in Ref. \cite{coles2012uncertainty}, where uncertainty relations in the presence of quantum memories are studied.

We define the states \mbox{$\rho_A=\D_A(\rho)$} and \mbox{$\rho_B=\D_B(\rho)$} to shorten the notation, and so have that
\begin{eqnarray*}
Q(A,\rho)+Q(B,\rho)&=&S(\rho||\rho_A)+S(\rho||\rho_B)\\
&\geq& S(\rho_B||D_B(\rho_A))+S(\rho||\rho_B),
\end{eqnarray*}
where the second line follows from the fact that the relative entropy is contractive under CP maps. Specializing to the projective/dephasing map we see that
\begin{eqnarray*}
S(\rho_B||\D_B(\rho_A))&=&-S(\rho_B)-\tr{\rho_B\ln(\D_B(\rho_A))}\\
&=&-S(\rho_B)-\tr{\rho \ln(\D_B(\rho_A)},
\end{eqnarray*}
which implies that
\begin{equation*}
S(\rho_B||\D_B(\rho_A))+S(\rho||\rho_B)=-S(\rho)-\tr{\rho \ln(\D_B(\rho_A))}.
\end{equation*}
This finally gives us\small 
\begin{subequations}
\begin{eqnarray}
Q(A,\rho)+Q(B,\rho)&\geq&-S(\rho)-\tr{\rho \ln(\D_B(\rho_A))},\label{eq:sumQ_1}\\
Q(A,\rho)+Q(B,\rho)&\geq&-S(\rho)-\tr{\rho \ln(\D_A(\rho_B))},\label{eq:sumQ_2}
\end{eqnarray}
\end{subequations}\normalsize
where the second inequality is obtained just by inverting the roles of $A$ and $B$. 

For the special case of the observables $A$ and $B$ being mutually unbiased, we have that \mbox{$\D_A(\rho_B) = \D_B(\rho_A) = \iden/d$}, which implies 
\begin{equation}
\label{eq:QEURMUB}
Q(A,\rho) + Q(B, \rho) \ge \ln d \left [  1- \frac{S(\rho)}{\ln d} \right ].
\end{equation}
For this, we find that we can factor out a ``purity" factor of $\left [ 1- \frac{S(\rho)}{\ln d}\right ]$ that accounts for the contribution from the classical uncertainty in the state. This bound turns out to be an optimal one. Let us also note that Eq. \eqref{eq:QEURMUB} implies a refinement of the Maassen-Uffink relation for mutually unbiased bases:
\be
H_A(\rho) + H_B(\rho) \geq \ln d + S(\rho),
\ee
which agrees with the result found in Ref. \cite{berta2010uncertainty} for the case of a trivial quantum memory.

\subsection{Purity-based bounds for arbitrary observables}

Beyond the case of mutually unbiased observables we see that the right-hand sides of inequalities Eqs. (\ref{eq:sumQ_1}) and (\ref{eq:sumQ_2}) may be written as
\begin{eqnarray*}
&&-S(\rho)-\sum_{i=1}^d{p_i(B,\rho)\ln\left(\sum_{j=1}^d |c_{ij}|^2p_j(A,\rho)\right)},\\
&&-S(\rho)-\sum_{i=1}^d{p_i(A,\rho)\ln\left(\sum_{j=1}^d |c_{ij}|^2p_j(B,\rho)\right)},
\end{eqnarray*}
where $c_{ij}=\braket{a_i}{b_j}$. One can obtain a convenient lower bound if one replaces the terms $|c_{ij}|$ by their maximum value $c_{AB}$, which provides the relation
\begin{equation}\label{nonMUB}
Q(A,\rho)+Q(B,\rho)\geq-2\ln c_{AB} \left [ 1 +\frac{S(\rho)}{2\ln c_{AB}} \right ].
\end{equation}
Let us again note that Eq. \eqref{nonMUB} implies a refinement of Maassen-Uffink uncertainty relation for arbitrary observables given by Eq. \eqref{eq:muEUR},
\be
\label{eq:improvedEUR}
H_A(\rho) + H_B(\rho) \geq -2 \ln c_{AB} + S(\rho).
\ee

The above derivation of Eq. \eqref{eq:improvedEUR} is much simpler than the one given in Ref. \cite{berta2010uncertainty}, which employs smooth entropies.  In addition, upon finishing this manuscript we became aware of the recent paper by Rudnicki \textit{et al.} \cite{rudnicki2014strong}, where majorization-based improvements over the term $-2 \ln c_{AB}$ have been proposed.

\subsection{Strong purity-based lower bounds}

In contrast to the MUB case, the bound given by Eq. \eqref{nonMUB} has the disadvantage of having a purity factor that is not independent from the observables $A$ and $B$. We might conjecture that a stronger bound holds, where the purity factor is independent of the observables, just like for the mutually unbiased observables [Eq. \eqref{eq:QEURMUB}],
\be
\label{eq:conjecture}
Q(A,\rho) + Q(B,\rho) \geq -2 \ln c_{AB} \left[ 1- \frac{S(\rho)}{\ln d}\right].
\ee
We see that this bound would be a linear SPB. Hence by Theorem \ref{th:nolinearbound}, it cannot hold for any dimension $d>2$. 

\begin{figure}[t!]
\includegraphics[width=\columnwidth]{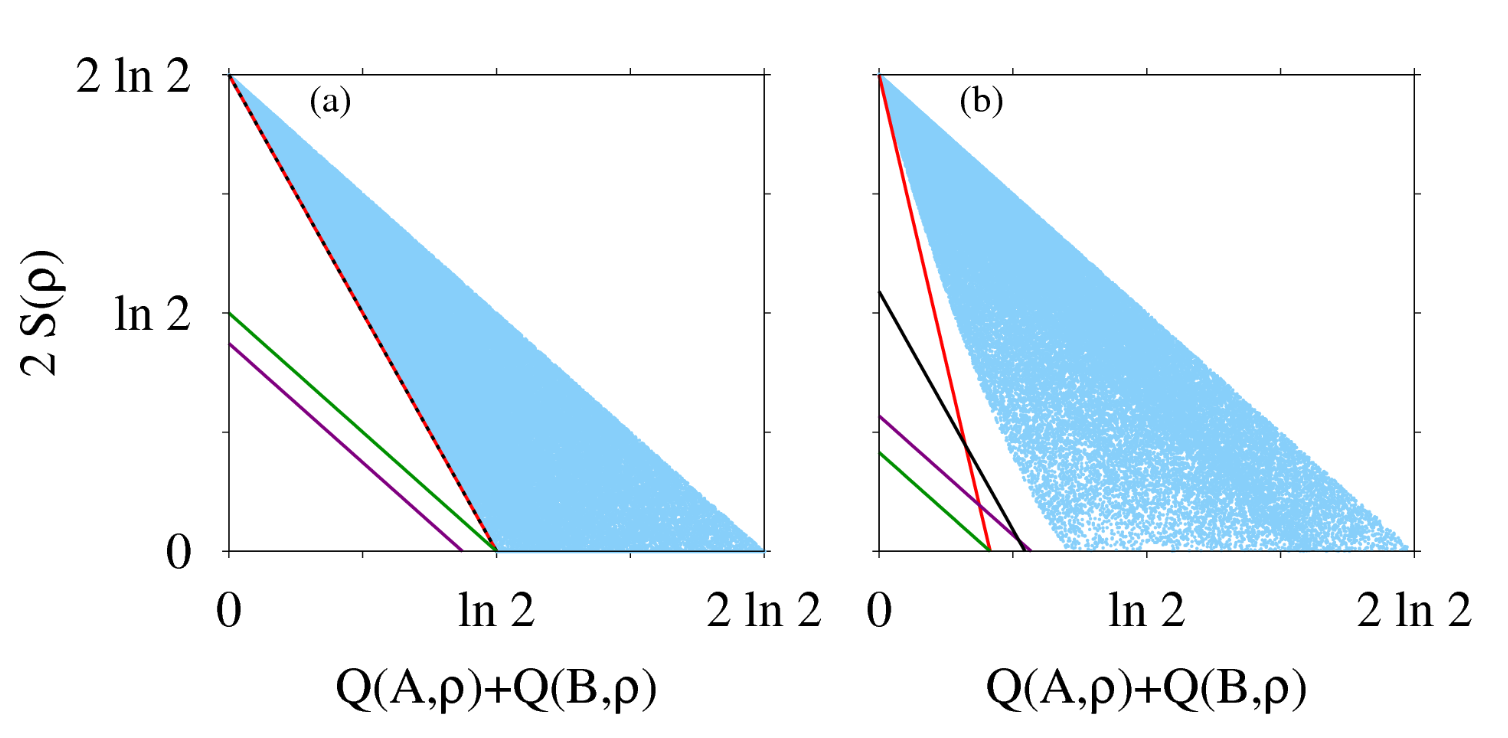}
\caption{\label{fig:qubit_bounds} Sum of classical vs. sum of quantum uncertainties plotted for $10^5$ random qubit states. The green line is the Maassen-Uffink bound. The red line is our strong purity-based bound. The black and purple lines are two of the recently proposed strong majorization bounds \cite{rudnicki2014strong}. The bounds are shown for (a) $A$ and $B$ mutually unbiased; (b) Eigenstates of $A$ and $B$ separated by angle $\gamma=\pi/3$ on the Bloch sphere, corresponding to $c_{AB}\approx 0.8660$.}
\end{figure}

The case $d=2$ is special, because an SPB exists and is given by
\be
\label{eq:strong_qubit}
Q(A,\rho) + Q(B,\rho) \geq -2 \ln c_{AB} \left[ 1- \frac{S(\rho)}{\ln 2}\right].
\ee
This qubit-specific uncertainty relation is proved in Appendix C by generalizing a proof of Tufarelli valid for the MUB case \cite{tufarelli2014}. In order to compare it with other known bounds we plot the sum of classical uncertainties, \mbox{$C(A,\rho) + C(B,\rho)=2S(\rho)$}, versus the sum of quantum uncertainties, \mbox{$Q(A,\rho)+Q(B,\rho)$}, for random quantum states $\rho$ (we will refer to such plots as QC uncertainty plots). In Fig. \ref{fig:qubit_bounds} this data is presented for qubit systems, together with our bound, the Maassen-Uffink bound and the strong majorization bounds of Ref. \cite{rudnicki2014strong}. As can be seen, for high purity states the majorization bounds outperform Eq. \eqref{eq:strong_qubit}, whereas for low purity states our bound outperforms the majorization bounds. None of the bounds are, however, optimal as they are linear in the von Neumann entropy, whereas numerics show that the minimum uncertainty curve in the \mbox{$Q(A,\cdot) + Q(B,\cdot)$} versus $2S(\cdot)$ plane is nonlinear (apart from the case of $A$ and $B$ being mutually unbiased). An interpolation between the majorization bounds of Ref. \cite{rudnicki2014strong} and our strong purity-based bound, Eq. \eqref{eq:strong_qubit}, appears to be the best currently available estimate of the total uncertainty valid for all $c_{AB}$ and all $S(\rho)$. 

Let us also note that the discussed relation for qubits implies, in terms of total uncertainties, the following strengthened version of the Maassen-Uffink bound:
\begin{equation}
H_A(\rho)+ H_B(\rho) \geq -2 \ln c_{AB} + 2 S(\rho) \left[1 + \frac{\ln c_{AB}}{\ln 2}\right].
\end{equation}

\section{Minimal uncertainty states with fixed purity}
\label{sec:section5}

As demonstrated, the purity-based bounds on the sum of quantum uncertainties are non-linear in the von Neumann entropy $S$. Therefore, in order to get more insight into their form, we will now focus directly on the states with fixed $S$ that minimize \mbox{$Q(A,\rho)+Q(B,\rho)$}. We will refer to them as minimal uncertainty states (MUS) with fixed purity (these are the states that form the optimal curve in the QC uncertainty plots). We restrict our considerations to the case of $d=2$ and show that even in this simplest scenario general MUS have non-trivial structure that is independent of the pure MUS with $S(\rho)=0$. Hence we will show that finding the optimal pure state is not enough to construct an optimal state with fixed $S>0$.

A general qubit observable has the form \mbox{$O=\alpha_1\iden+\alpha_2\v{o}\cdot\v{\sigma}$}, where $\v{\sigma}$ denotes the vector of Pauli operators and $\v{o}$ is the Bloch vector. However, as entropic uncertainty measures depend only on the eigenstates of observables and not on their eigenvalues, we can restrict our considerations to observables \mbox{$A=\v{a}\cdot\v{\sigma}$} and \mbox{$B=\v{b}\cdot\v{\sigma}$}. Without the loss of generality one can choose 
\begin{eqnarray*}
\v{a}&=&(0,0,1),\\
\v{b}&=&(\sin\gamma,0,\cos\gamma),
\end{eqnarray*}
and it is enough to restrict to $\gamma\in[0,\pi/2]$, as for entropic quantities $\v{a}$ and $-\v{a}$ are indistinguishable. In this setting one has $c_{AB}=\cos\gamma$. 

The form of minimal uncertainty pure states for qubits has been studied previously \cite{sanches1998optimal,ghirardi2003optimal} and shown to exhibit the following dependence on $\gamma$. For $\gamma<\gamma_c$ (where \mbox{$\gamma_c\approx 1.17056$} was found numerically, see Ref. \cite{ghirardi2003optimal} for details), the optimal state is represented by the Bloch vector 
\be
\v{c}_<=(\sin\gamma/2,0,\cos\gamma/2);
\ee
i.e., it lies in the middle between the eigenstates of $A$ and $B$ on the Bloch sphere. For $\gamma>\gamma_c$ a parametric bifurcation occurs -- the number of optimal states doubles and they are represented by the Bloch vectors
\be
\v{c}_>=(\sin\gamma/2\pm\beta,0,\cos\gamma/2\pm\beta),
\ee
where $\beta$ is a non-elementary function of $\gamma$ that increases from $\beta=0$ for $\gamma=\gamma_c$ to $\beta=\gamma/2$ for $\gamma=\pi/2$. In the Bloch sphere picture when $\gamma$ exceeds $\gamma_c$ the two optimal states start to move away from the vector lying symmetrically between the eigenstates of $A$ and $B$, and move toward these eigenstates, eventually overlapping with them for $\gamma=\pi/2$, i.e., for $A$ and $B$ being mutually unbiased.

\begin{figure}[t!]
\includegraphics[width=\columnwidth]{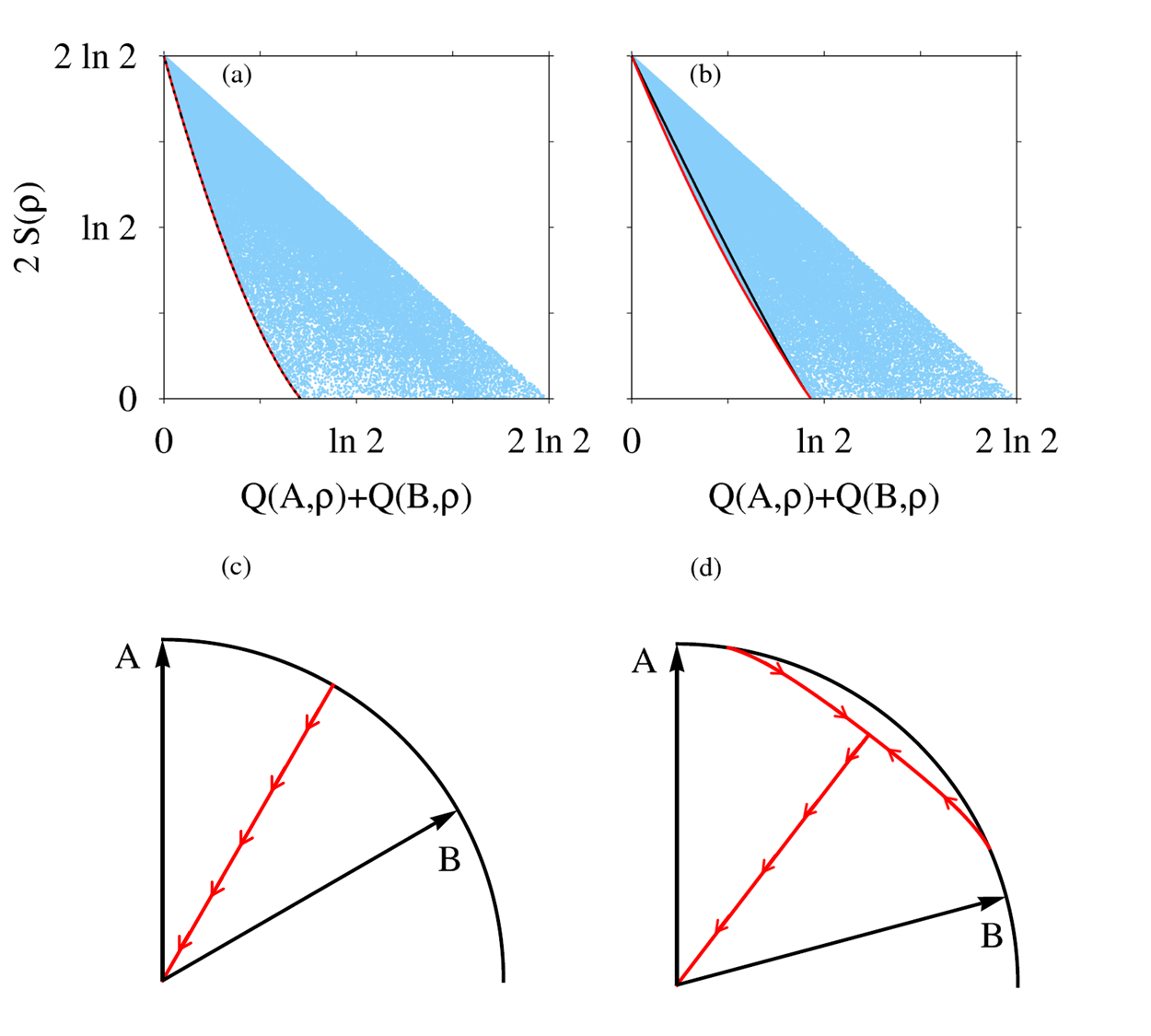}
\caption{\label{fig:qubit_MUS}(a, b) Sum of classical vs. sum of quantum uncertainties plotted for $10^5$ random qubit states. The red line represents the numerically optimized MUS with fixed purity, whereas the black line represents the mixture of pure MUS with maximally mixed state. In (a) the eigenstates of $A$ and $B$ are separated by \mbox{$\gamma=60^o<\gamma_c$}, in (b) the eigenstates of $A$ and $B$ separated by \mbox{$\gamma=75^o>\gamma_c$}.  (c, d) The trajectory, parametrized by $S \in [0,\ln 2]$, of MUS states with fixed purity in the first quadrant of the plane spanned by $\v{a}$ and $\v{b}$. In (c) $A$ and $B$ are given as in (a), in (d) they are given as in (b).}
\end{figure}

\begin{figure}[t!]
\includegraphics[width=\columnwidth]{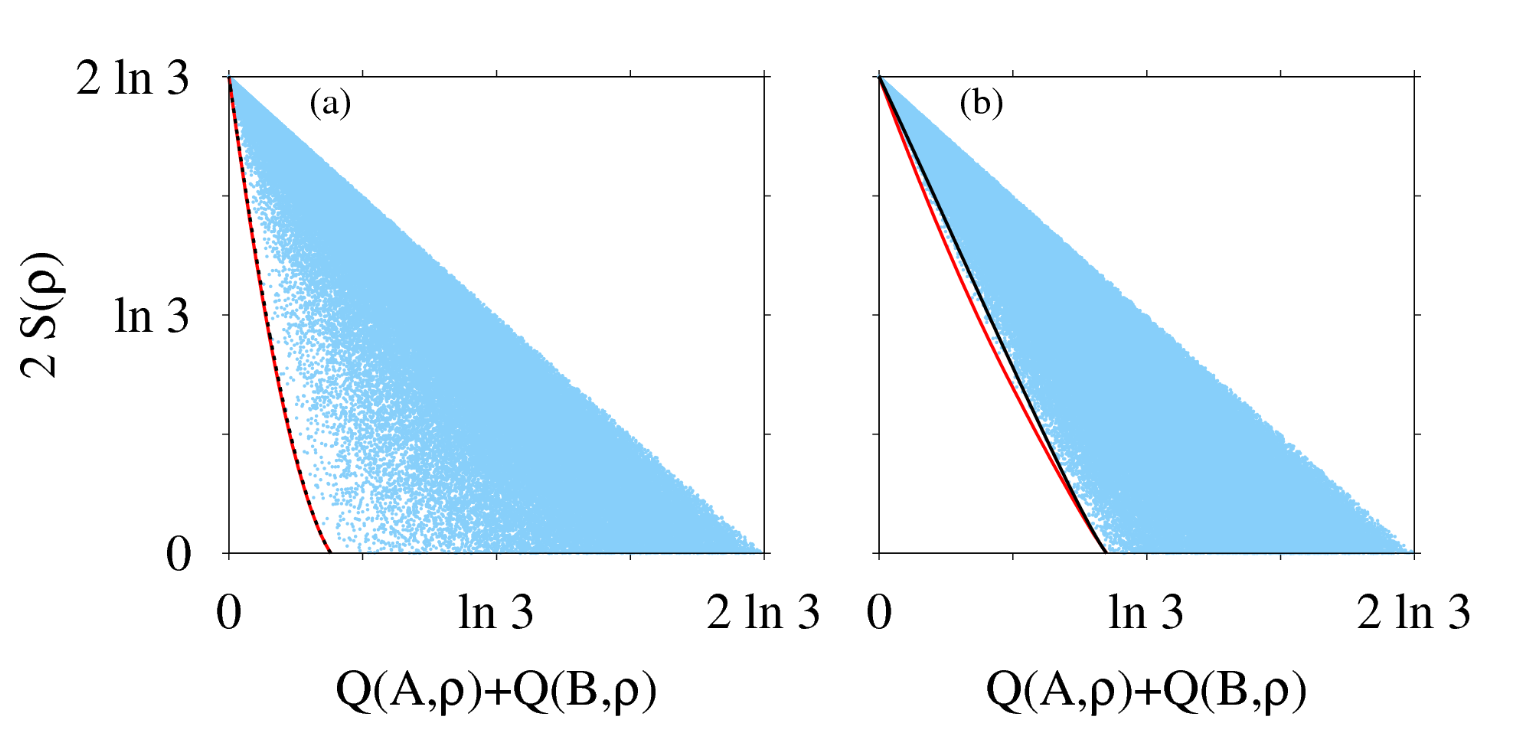}
\caption{\label{fig:qutrit_MUS} Sum of classical vs. sum of quantum uncertainties plotted for $10^5$ random qutrit states. The red line represents the numerically optimized MUS with fixed purity, whereas the black line represents the mixture of pure MUS with maximally mixed state. Observables $A$ and $B$ chosen so that the eigenstates of $B$ are connected with the eigenstates of $A$ by rotation around (1,1,1) axis by angle (a) $\alpha=\pi/6$, corresponding to $c_{AB}\approx 0.9107$ and (b) $\alpha=\pi/3$, corresponding to $c_{AB}\approx 0.6667$.}
\end{figure}

One might suspect that the general MUS with fixed purity can be obtained just by mixing the pure MUS with the maximally mixed state. Interestingly however, we will show that this is not the case, which supports the claim that MUS with fixed purity are not just a trivial extension of pure MUS. The behavior of qubit MUS with fixed purity is shown in Fig. \ref{fig:qubit_MUS}. It is easy to see that a qubit MUS, independently of their purity, must lie in the plane spanned by $\v{a}$ and $\v{b}$, thus having the form
\be
\v{c}=r(\sin\theta,0,\cos\theta),
\ee
where $r\in[0,1]$. A direct inspection shows that for $\gamma<\gamma_c$, when a pure MUS is represented by the Bloch vector $\v{c}_<$, the general MUS with fixed purity is given by $r\v{c}_<$, i.e., the same Bloch vector, just shorter [see Figs. \ref{fig:qubit_MUS}(a) and \ref{fig:qubit_MUS}(c)]. Hence in this regime the general MUS is obtained by mixing the pure MUS with maximally mixed state. However, for $\gamma>\gamma_c$ this is no longer the case, as for a given $\v{c}_>$ decreasing the purity decreases $\beta$, and the MUS states flow toward the $r \v{c}_<$ solution [see Figs. \ref{fig:qubit_MUS}(b) and \ref{fig:qubit_MUS}(d)]. Numerical investigations performed for qutrits suggests that this nontrivial structure of the MUS is a general feature, not only limited to qubits; see Fig.~\ref{fig:qutrit_MUS}.

Finally, we would also like to make a short comment on the states that are the opposite of MUS with fixed purity -- the maximal uncertainty states with fixed purity. As can be seen in Figs. \ref{fig:qubit_bounds}--\ref{fig:qutrit_MUS}, these states form a straight line in QC uncertainty plots, connecting maximally mixed state and the pure state $\ket{\psi^*}$, which is unbiased in eigenbases of both observables (see Lemma 1). Thus, the states of fixed purity that maximize the sum of quantum uncertainties have a particularly simple form: \mbox{$p\iden/d+(1-p)\ketbra{\psi^*}{\psi^*}$}. Note that this means that for every fixed purity there exists a state that maximizes the sum of total uncertainties, i.e., for which \mbox{$H_A+H_B=2\log d$}.

\section{Discussion and outlook}

Entropic uncertainty relations are powerful relations that capture the inevitable trade-off in our ability to prepare a quantum system in a state that has highly peaked distributions for two non-commuting observables. Obviously the issue of ``classical noise" arises due to imperfect experimental preparations. However, a similar scenario arises if we prepare a pure entangled state and are interested in the uncertainties of two observables $A$ and $B$ on a particular subsystem. How should we then cleanly separate the uncertainty that arises due to the intrinsic noncommutativity of the observables from that which does not? In this work, we have proposed a natural decomposition of the total uncertainties into measures that respect basic desiderata one would require. We showed, very generally, that any measure of total quantum uncertainty will not have state-independent constraints but instead could be bounded relative to the degree of purity present in the state. The particular measure of quantum uncertainty in a state we consider is $S(\rho || \D_A(\rho)) + S(\rho || \D_B (\rho))$, namely the sum of the relative entropies to the state dephased in the eigenbases of $A$ and $B$ and the measure of purity is taken to be $S(\rho)$. The entropic decomposition into classical and quantum components leads to ``purity-based bounds.'' We moved the first steps toward a study of the highly nontrivial structure of minimal uncertainty states in the presence of classical uncertainty and we presented a general result about the nonlinearity of all purity-based bounds on the sum of quantum uncertainties that are at least as good as Maassen-Uffink relation and that reduce to zero for the maximally mixed (classical) state. These bounds provide generalizations to the traditional entropic relations, stressing the role that classical noise plays into them and opening up a new direction that remains largely unexplored.

\textbf{Acknowledgements:} We would like to thank T. Tuffarelli and G. McConnell for very useful discussions as well as \L{}. Rudnicki and P. Coles for their comments on the first version of our manuscript. This work was supported by EPSRC. D.J. is supported by the Royal Society. T.R. is supported by the Leverhulme Trust.

\bibliographystyle{prsty}
\bibliography{bibliographyabb}
\appendix

\section{Extension to general projective measurements}

\subsection{Classical-quantum uncertainty splitting}

The classical-quantum uncertainty splitting can be extended to general degenerate observables. The corresponding projective measurement is no longer sharp (rank-1), and so it is natural to demand that the classical uncertainty should reflect this degeneracy.
Intuitively, the more a measurement coarse-grains the Hilbert space, the smaller the classical uncertainty will be.

Consider a state $\rho$ and a projective measurement $\{\Pi_i\}$. Define
\be
\nonumber
\rho_i = \frac{\Pi_i \rho \Pi_i}{p_i},
\ee
where $p_i = \mathrm{Tr}[\Pi_i \rho \Pi_i]$. Let us also define the measurement map
\be
\nonumber
\rho \longmapsto D_{\Pi}(\rho) = \sum_i p_i \rho_i = \sum_i \Pi_i \rho \Pi_i,
\ee
which associates to each state the post-measurement state (without post-selection).
The relative entropy between the initial state and the post-measurement state is given by
\begin{eqnarray}
S(\rho||D_{\Pi}(\rho)) & = & -S(\rho) - \mbox{Tr}[\rho \ln D_{\Pi}(\rho)] \nonumber \\
& = & -S(\rho) - \mbox{Tr}[D_{\Pi}(\rho) \ln D_{\Pi}(\rho)] \nonumber \\
& = & -S(\rho) + S(D_{\Pi}(\rho)).
\label{eq:gensplit1}
\end{eqnarray}
Given that $\{\rho_i\}$ have orthogonal support one has
\be
\nonumber
 S(D_{\Pi}(\rho)) = S\left(\sum_i p_i \rho_i \right) = H_{\Pi}(\rho) + \sum_i p_i S(\rho_i).
\ee
Substituting this in Eq. \eqref{eq:gensplit1} one obtains
\be
\nonumber
S(\rho||D_{\Pi}(\rho)) = -S(\rho) + H_{\Pi}(\rho) + \sum_i p_i S(\rho_i),
\ee
which gives the final splitting
\be
\label{eq:gensplitting}
H_{\Pi}(\rho) = Q_{\Pi}(\rho) + C_{\Pi}(\rho),
\ee
with
\begin{subequations}
\begin{eqnarray}
Q_{\Pi}(\rho) &=& S(\rho||D_{\Pi}(\rho)),\\
C_{\Pi}(\rho) &=& S(\rho) - \sum_i p_i S(\rho_i).\label{eq:classical}
\end{eqnarray}
\end{subequations}
Notice that the classical uncertainty is now a function of the measurement $\{\Pi_i\}$. This is because the uncertainty depends on degeneracy of the measured observable, so on the coarse-graining (sharpness) of the corresponding measurement and in general it will be lower than the von Neumann entropy, which is the uncertainty for a perfectly discriminating measurement. Consider, for example, the qutrit state
\be\nonumber
\xi = \frac{1}{2}(\ket{0}\bra{0} + \ket{1}\bra{1}),
\ee
and the projective measurement
\be\nonumber
\Pi_1 = \ket{0}\bra{0} + \ket{1}\bra{1}, \quad\Pi_2 = \ket{2}\bra{2}.
\ee
Even though the von Neumann entropy of $\xi$ is nonzero, the classical uncertainty of such $\{\Pi_i\}$ measurement on $\xi$ should vanish (as outcome associated with $\Pi_1$ has probability 1) and, as can be easily checked with the definition given by Eq. \eqref{eq:classical}, it does vanish.

\subsection{Luo's axioms for degenerate observables}

Let us check that the introduced quantities satisfy the basic criteria we imposed. We add a proof when the property is not a trivial extension of the nondegenerate case.
\begin{enumerate}

\item $C_{\Pi}$ vanishes on pure states.

\item $Q_{\Pi}(\rho) = 0$ if and only if the measurement is classical; i.e., $[\rho, \Pi_i] = 0$ \; $\forall i$. This generalizes Luo axiom~\ref{luo2}.

\item $Q_{\Pi}$ is a convex function as it is defined in terms of relative entropy. To show that $C_{\Pi}$ is a concave function simply note that $H_{\Pi}$ is a concave function and taking into account the additive splitting, Eq. \eqref{eq:gensplitting}, together with convexity of $Q_{\Pi}$, it is easy to prove that $C_{\Pi}$ must be concave.

\item $0\leq C_{\Pi}(\rho) \leq H_{\Pi}(\rho)$: it suffices to observe that our $C_{\Pi}(\rho)$ is equal to the QC-mutual information introduced by Sagawa \cite{sagawa2008second} and independently by Groenewold \cite{groenewold1971problem} and Ozawa \cite{ozawa1986information}. This quantity is known to satisfy the bounds given. Our measure of classical uncertainty $C_{\Pi}$ enters the second law of thermodynamics with feedback control, measuring the amount of extra work that can be extracted from a system using feedback \cite{sagawa2008second, sagawa2012thermodynamics}.

Another property that is not among Luo axioms, but that supports our interpretation of $C_{\Pi}$ is the following (for a proof, see Ref. \cite{sagawa2012thermodynamics}). If a measurement $\Pi'$ is a refinement of a measurement $\Pi$, then
\be
\forall \rho\quad C_{\Pi}(\rho)\leq C_{\Pi'}(\rho).
\ee
In other words, $C_{\Pi}(\rho)$ decreases under coarse-graining, as expected.
\end{enumerate}

\section{No linear SPB theorem}

The proof proceeds in three steps: first, we construct the weakest linear bound satisfying desiderata \ref{desiderata1} and \ref{desiderata2}. A violation of this bound will imply the violation of any other bound with the same properties. Second, we produce counterexamples for dimension $3,4,5$, i.e., construct observables $A$ and $B$ as well as states $\rho_3$, $\rho_4$, and $\rho_5$ for which the weakest bound is violated. Third, we show that these already imply that a counterexample exists in any dimension $d>5$.

\bigskip

\emph{Step 1}. All bounds linear in $S$ will have the form
\be
f(A,B,S(\rho)) = a + b S(\rho),
\ee
for some $a, b \in \mathbb{R}$. Assumption \ref{desiderata2} implies
\be
f(A,B, \ln d) = a + b \ln d= 0 \Rightarrow b= -a/\ln d,
\ee
hence 
\be
f(A,B, S(\rho)) = a \left( 1 - \frac{S(\rho)}{\ln d}\right).
\ee
Assumption \ref{desiderata1} implies $a \geq -2\ln c_{AB}$, 
\be
\label{eq:genboundbound}
f(A,B, S(\rho)) \geq -2 \ln c_{AB} \left( 1 - \frac{S(\rho)}{\ln d}\right):=f_{w}(A,B, S(\rho)),
\ee
where $f_w$ denotes the weakest linear bound satisfying the requirements \ref{desiderata1} and \ref{desiderata2}. Notice that for any $d >2$,
\be
\label{eq:genboundbound2}
f_{w}(A,B, S(\rho)) \geq -2 \ln c_{AB} \left( 1 - \frac{S(\rho)}{\ln 3}\right).
\ee

\bigskip

\emph{Step 2}. Let $\ket{a_i}$ and $\ket{b_i}$ denote the eigenstates of observables $A$ and $B$ acting on a $d$-dimensional Hilbert space. Assume that these two bases are linked by a rotation $R_d = \exp(\theta_d S_d)$, where $S_d$ is the $d$-dimensional skew-symmetric matrix such that $|S_{ij}|= 1- \delta_{ij}$ and $\theta_d$ are real numbers. We choose
\be
\theta_3 = 4\pi/7, \quad \theta_4= \pi/2, \quad \theta_5 = \pi.
\ee

The following states, written in the $\ket{a_i}$ basis, violate the bound of Eq. \eqref{eq:genboundbound} in dimensions \mbox{$d=3,4,5$}, respectively:
\be
\nonumber
\rho_3= \left[ \begin{matrix}
  0.61  &&  -0.15 &&      0 \\
   -0.15  &&  0.26  &&  0.08 \\
       0   && 0.08 &&   0.13
\end{matrix} \right],
\ee
\be
\nonumber
\rho_4= \left[ \begin{matrix}
 0.08  &&  0.03  &&  0.04 &&  -0.03 \\
     0.03  &&  0.06 &&  -0.03  &&  0 \\
     0.04 &&  -0.03  &&  0.08 &&  -0.03 \\
    -0.03 &&   0  &&    -0.03  &&  0.78
\end{matrix} \right],
\ee
\be
\nonumber
\rho_5= \left[ \begin{matrix}
  0.19  &&  0.05  && -0.06  && -0.02 &&  -0.05 \\
    0.05 &&   0.49 &&  -0.11   &&    0  &&     0 \\
   -0.05 &&  -0.11 &&   0.22 &&  -0.03  &&  0.02 \\
   -0.02 &&      0 &&  -0.03  &&  0.05  &&     0 \\
   -0.05  &&     0 &&   0.02   &&    0  &&  0.05
\end{matrix} \right].
\ee

\bigskip

\emph{Step 3}. We will now show that the existence of the counterexample in dimension $3$ immediately implies the existence of counterexamples in any dimension $d > 5$. Hence our counterexamples for $d=3,4,5$ immediately imply the result for all $d >2$. 

Fix a $d$-dimensional Hilbert space $\mathcal{H}$ and consider the $3$-dimensional subspace $\mathcal{H}_3$ spanned by $\{\ket{a_i}\}$ and $\{\ket{b_i}\}$ given by the counterexample of Step 2. We can complete $\{\ket{a_i}\}$ and $\{\ket{b_i}\}$ to a basis in \mbox{$\mathcal{H}=\mathcal{H}_3\oplus\mathcal{H}_3^{\perp}$} by choosing 
\be
\braket{a_i}{b_j} = \frac{1}{\sqrt{d-3}}, \; \; \forall i,j =4,...,d.
\ee
This can be done as $\mathcal{H}_3^{\perp}$ is a $d-3$-dimensional Hilbert space and one can always find two mutually unbiased bases as long as $d>4$. Let us call \mbox{$c^{(3)}_{AB} = \max_{i,j=1}^3 |\braket{a_i}{b_j}|=0.6851$}. Notice that 
\be
\nonumber
c^{(d)}_{AB} := \max_{i,j=1}^d |\braket{a_i}{b_j}| = \max \{c^{(3)}_{AB}, 1/\sqrt{d-3}\}=c^{(3)}_{AB},
\ee
for all $d\geq 6$. Exploiting this construction and Eq. \eqref{eq:genboundbound2} it is easy to see that $\rho_3$ (seen now as a quantum state in $\mathcal{H}$) violates the bound of Eq. \eqref{eq:genboundbound} in any dimension $d >5$.

\section{Strong-purity based bound for qubit}

\begin{figure}[h!]
\includegraphics[width=0.7\columnwidth]{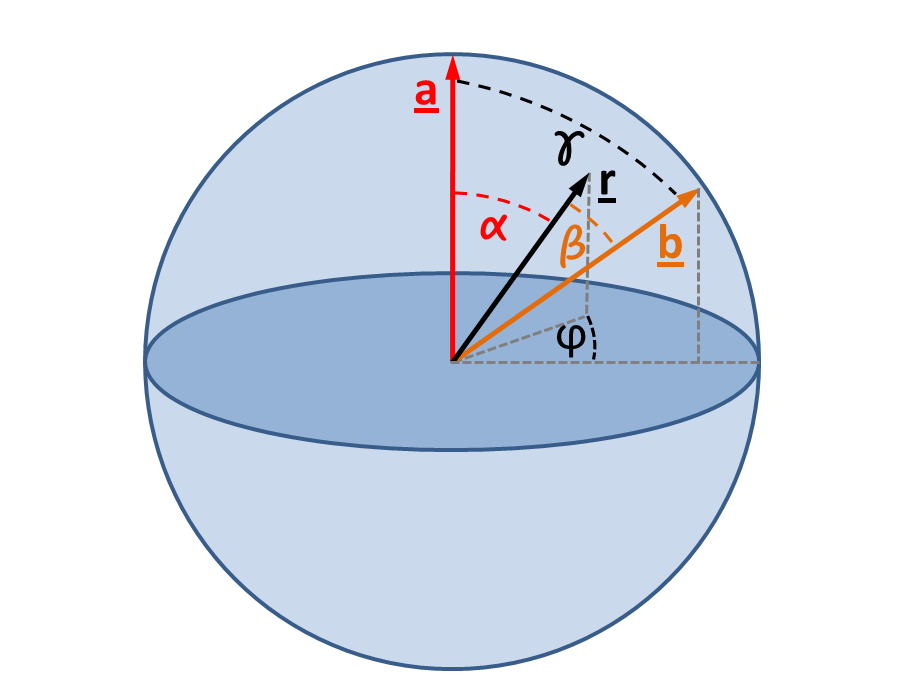}
\caption{\label{fig:bloch_vectors}The orientation of the Bloch sphere.}
\end{figure}

We prove Eq. \eqref{eq:conjecture} in the case $d=2$. As explained in Sec. \ref{sec:section5}, without loss of generality we can restrict our considerations to two single qubit observables given by $A=\v{a}\cdot\v{\sigma}$, and $B=\v{b}\cdot\v{\sigma}$, where $\v{\sigma}=(X,Y,Z)$ denotes the vector of Pauli matrices and $\v{a}$ and $\v{b}$ are the Bloch vectors. Without loss of generality we may assume that $\v{a}=(0,0,1)$ and $\v{b}=(\sin\gamma,0,\cos\gamma)$,
where $\gamma\in[0,\pi/2]$ (extending the range over $\pi/2$ is unnecessary, as for the entropic quantities $\v{a}$ and $-\v{a}$ are indistinguishable). A general qubit state can now be written as 
\be
\rho = \frac{\iden + \v{r} \cdot \v{\sigma}}{2},
\ee
with
\be
\v{r}= r (\sin \alpha \cos \varphi, \sin \alpha \sin \varphi, \cos \alpha),
\ee
$r\in[0,1]$, $\alpha\in[0,\pi]$, and $\varphi\in[0,2\pi]$. All three Bloch vectors $\v{a}$, $\v{b}$, and $\v{r}$ are depicted on the Bloch sphere in Fig. \ref{fig:bloch_vectors}. The probability distributions of the outcomes of $A$ and $B$ in a state $\rho$ are given by
\begin{subequations}
\label{eq:probdistribution}
\begin{eqnarray}
p_A&=&\left( \frac{1+ r \cos \alpha}{2}, \frac{1- r \cos \alpha}{2}\right), \\
p_B&=&\left( \frac{1+ r \cos \beta}{2}, \frac{1- r \cos \beta}{2}\right),
\end{eqnarray}
\end{subequations}
where $\cos\beta=\cos \varphi \sin \alpha \sin \gamma+\cos\alpha\cos\gamma$.
Introducing the binary entropy,
\be
\nonumber
H_2(p)=-p \ln p - (1-p)\ln(1-p),
\ee
the improved entropic uncertainty relation, Eq. \eqref{eq:conjecture}, for a qubit reduces to\footnotesize
\begin{eqnarray}
\label{eq:entropic_qubit}
&&H_2\left( \frac{1+ r \cos \alpha}{2}\right)+H_2\left( \frac{1 + r \cos\beta}{2}\right)+2\ln\left(\cos\frac{\gamma}{2}\right)\nonumber\\
&&-2H_2\left(\frac{1+r}{2}\right)\left(1+\frac{\ln\left(\cos\frac{\gamma}{2}\right)}{\ln 2}\right)\geq 0.
\end{eqnarray}
\normalsize In order to prove that the above inequality always holds let us first denote its left-hand side by \mbox{$F(\alpha,\beta,\gamma,r)$}. We will find its minimum and show that it is greater or equal to zero. Computing the derivative with respect to $r$,
\begin{eqnarray*}
\frac{\partial}{\partial r}F(\alpha,\beta,\gamma,r)&=&-\tanh^{-1}(r\cos\alpha)\cos\alpha\\
&&-\tanh^{-1}(r\cos\beta)\cos\beta\\
&&+2\tanh^{-1}(r)\left(1+\frac{\ln\left(\cos\frac{\gamma}{2}\right)}{\ln 2}\right),
\end{eqnarray*}
we will show this is always nonnegative. Using the Taylor series representation of $\tanh^{-1}$ one finds that the above derivative is given by
\small
\be
\nonumber
\sum_{n=0}^{\infty}\frac{r^{2n+1}}{2n+1}\left(2-\cos^{2n+2}\alpha-\cos^{2n+2}\beta+\frac{2}{\ln 2}\ln\left(\cos\frac{\gamma}{2}\right)\right).
\ee
\normalsize
Now let us denote the coefficient in parentheses standing by the $n$-th term by $a_n$ and note that
\be
\nonumber
\forall n\quad a_0\leq a_n.
\ee
Therefore, if we can show that $a_0\geq 0$ then all the coefficients are positive and, taking into account the positivity of $r$, the considered derivative is positive for all $\alpha$, $\beta$, and $\gamma$. It is easy to see, using the explicit dependence of $\cos\beta$ on $\alpha$, $\gamma$, and $\varphi$, that
\be
a_0\geq \sin^2\alpha+\frac{2}{\ln 2}\ln\left(\cos\frac{\gamma}{2}\right)+\min_{+,-}\sin^2(\alpha\pm\gamma).
\ee
Introducing \mbox{$\delta=\alpha \pm \gamma/2$} one gets that for $a_0\geq 0$ to be true one has to prove that
\be
1-\cos 2\delta\cos\gamma+\frac{2}{\ln 2}\ln\left(\cos\frac{\gamma}{2}\right) \geq 0.
\ee
The minimum of the left-hand side of the above inequality is achieved for $\delta=0$, hence it is enough to prove that
\be
\nonumber
2\sin^2\frac{\gamma}{2}+\frac{2}{\ln 2}\ln\left(\cos\frac{\gamma}{2}\right) \geq 0,
\ee
which can be easily done. Therefore, we have proven that 
\be
\forall \alpha,\beta,\gamma\quad\frac{\partial}{\partial r}F(\alpha,\beta,\gamma,r)\geq 0,
\ee
and it is easy to see that
\be
\forall \alpha,\beta,\gamma\quad F(\alpha,\beta,\gamma,0)=0.
\ee
Hence,
\be
\forall \alpha,\beta,\gamma,r\quad F(\alpha,\beta,\gamma,r)\geq 0,
\ee
so that the conjectured bound for qubit holds.

\end{document}